\documentclass[11pt]{article}
\usepackage[utf8]{inputenc}

\usepackage[margin=1in]{geometry}

\usepackage{xcolor}
\title{Draft notes}
\date{}

\usepackage[english]{babel}
\usepackage{cite}
\usepackage{mcite}
\usepackage{graphicx}
\usepackage{amsmath,amssymb,amsfonts,mathrsfs}
\usepackage{shuffle}
\usepackage{ifdraft}

\allowdisplaybreaks[4]

\DeclareMathOperator*{\Reglim}{Reg}

\newcommand{\Hf}[5]{ 
 {}_{#1}F_{#2}  \left (\begin{smallmatrix}
   {#3} \\
   {#4}
\end{smallmatrix}\Bigr|#5\right)
  }

\newcommand{\col}[2]{ 
    \left (\begin{smallmatrix}
   {#1} \\
   {#2}
\end{smallmatrix}\right)
  }

\newcommand{\Ef}[3]{ 
    E_4  \left (\begin{smallmatrix}
   {#1} \\
   {#2}
\end{smallmatrix};#3\right)
  }

\newcommand{\Eff}[5]{ 
    E_4  \left (\begin{smallmatrix}
   {#1} & {#3} \\
   {#2} & {#4}
\end{smallmatrix};#5\right)}

\newcommand{\Efff}[7]{ 
    E_4  \left (\begin{smallmatrix}
   {#1} & {#3} & {#5}\\
   {#2} & {#4} & {#6}
\end{smallmatrix};#7\right)}

\newcommand{\Effff}[9]{ 
    E_4  \left (\begin{smallmatrix}
   {#1} & {#3} & {#5} & {#7}\\
   {#2} & {#4} & {#6} & {#8}
\end{smallmatrix};#9\right)}

\newcommand{\ep}{\varepsilon}

\newcommand{\bea}{\begin{eqnarray}}
\newcommand{\eea}{\end{eqnarray}}
\newcommand{\be}{\begin{equation}}
\newcommand{\ee}{\end{equation}}

\begin{document}

\begin{titlepage}
\renewcommand{\thefootnote}{\fnsymbol{footnote}}

\begin{center}
{\Large \bf Sunrise integral with two internal masses and pseudo-threshold kinematics in terms of elliptic polylogarithms}
\end{center}

\par \vspace{2mm}
\begin{center}
{\sc J.~Campert $^{(a)}$, F.Moriello $^{(a)}$, A.~Kotikov$^{(b)}$}
\vspace{5mm}

$^{(a)}$ {\normalsize\it
  ETH Z\"urich, Institut f\"ur theoretische Physik, Wolfgang-Pauli str. 27,}\\
{\normalsize\it 8093, Zürich, Switzerland}\\

$^{(b)}${\normalsize\it Bogolyubov Laboratory for Theoretical Physics, JINR,}\\
{\normalsize\it 141980 Dubna (Moscow region), Russia} \\

\vspace{5mm}

\end{center}

\par \vspace{2mm}
\begin{center} {\large \bf Abstract} \end{center}
\begin{quote}
\pretolerance 10000

We consider a two-loop sunrise integral with two different internal masses at pseudo-threshold kinematics and we solve it in terms of elliptic polylogarithms to all orders of the dimensional regulator.

\end{quote}

\vspace*{\fill}

\end{titlepage}

\tableofcontents

\section{Introduction}
\label{sec:intro}

In the last decades much progress has been made in the understanding of the mathematical properties of Feynman integrals. Arguably many of the breakthroughs in this line of research originated from the identification of classes of special functions suited for the solution of Feynman integrals by means of various analytic methods. It is a well-known fact that while many Feynman integrals admit representations in terms of so-called multiple polylogarithms (MPLs) \cite{Goncharov:1998kja,Remiddi:1999ew}, this space of functions is not sufficient to express integrals when the number of physical scales is sufficiently large. More recently, the scientific community has centered its attention to the study of Feynman integrals whose geometric properties are defined by elliptic curves. Following early investigations of \cite{Sabry:1962} and \cite{Broadhurst:1993mw}, many integrals involving elliptic curves have been computed in the literature \cite{Laporta:2004rb,Kniehl:2005bc,Bailey:2008ib,MullerStach:2011ru,
  Adams:2013kgc,Bloch:2013tra,Remiddi:2013joa,Adams:2014vja,Adams:2015gva,Adams:2015ydq,Bloch:2016izu,Adams:2017ejb,Bogner:2017vim,Adams:2018yfj,Honemann:2018mrb,Bloch:2014qca,Sogaard:2014jla,Tancredi:2015pta,Primo:2016ebd,Remiddi:2016gno,Adams:2016xah,Bonciani:2016qxi,vonManteuffel:2017hms,Adams:2017tga,Ablinger:2017bjx,Primo:2017ipr,Passarino:2016zcd,Remiddi:2017har,Bourjaily:2017bsb,Hidding:2017jkk,Broedel:2017kkb,Broedel:2017siw,Broedel:2018iwv,Lee:2017qql,Lee:2018ojn,Adams:2018bsn,Adams:2018kez,Broedel:2018qkq,Bourjaily:2018yfy,Bourjaily:2018aeq,Besier:2018jen,Mastrolia:2018uzb,Ablinger:2018zwz,Frellesvig:2019kgj,Frellesvig:2019byn,Bonciani:2019jyb,Francesco:2019yqt,Broedel:2019hyg,Blumlein:2019svg,Broedel:2019tlz,Bogner:2019lfa,Kniehl:2019vwr,Broedel:2019kmn,Abreu:2019fgk,Duhr:2019rrs,Klemm:2019dbm,Bonisch:2020qmm,Walden:2020odh}.

In a parallel line of research, a class of functions, the so-called
Elliptic Multiple Polylogarithms, describing all iterated integrals on the torus has been identified in the mathematics literature \cite{Brown:2011} (see also \cite{Beilinson:1994,Levin:2007}). While these functions formally solve the problem of generalising MPLs to more complicated geometries, their definition is not naturally suited for physical applications. Progress in this direction has been made in \cite{Broedel:2017kkb}, where eMPLs are defined on the complex plane, and their structure naturally adapts to representations of Feynman integrals commonly used in the physics literature (e.g. Feynman parameters).

Special functions such as MPLs and eMPLs, frequently appear when computing Feynman integrals in dimensional regularisation. More specifically, Feynman integrals admit a Laurent expansion with respect to the dimensional regulator and the coefficients of this expansion can be often computed explicitly in terms of known special functions. In practice it is often possible to truncate the Laurent series, as the computation of physically relevant quantities requires only a few expansion orders. Nonetheless it is interesting to explore the structure of these coefficients at higher orders or, more generally, to all orders of the dimensional regulator. In this context, all orders results for the equal mass sunrise integral have been obtained in \cite{Adams:2017ejb} in terms of iterated integrals of modular forms.

In this paper we consider a two-loop sunrise integral topology with two internal masses and pseudo-threshold kinematics \cite{Kalmykov:2008ge} (see also \cite{Kniehl:2005bc,Kniehl:2019vwr}). More precisely, we consider two different internal masses, denoted by $m$ and $M$, and external kinematics $p^2=-m^2$. This integral family appears when considering non-relativistic limits of Quantum Chromodynamics (NRQCD) and Quantum Electrodynamics (NRQED) (see for example \cite{Kniehl:2005bc} and \cite{Kniehl:2019vwr}). Examples of phenomenological applications of NRQCD are the study of heavy-quarkonium production and decay
(see Refs.
\cite{Sang:2015uxg,Chen:2017xqd,Feng:2017hlu,Chen:2017pyi,Chen:2017soz,Yang:2020pyh}
and references therein) and the near-threshold production of $t\bar{t}$
\cite{Adams:2018bsn,Adams:2018kez,Bonciani:2009nb,Czakon:2009zw,Bonciani:2010mn,Beneke:2011mq,Cacciari:2011hy,Czakon:2013goa,Bonciani:2013ywa,Beneke:2015kwa,Czakon:2015owf,Czakon:2017wor,Beneke:2017rdn,Chen:2019zoy,Becchetti:2019tjy,DiVita:2019lpl,Bonciani:2019jsw,Cooper-Sarkar:2020twv,Bonciani:2020tvf}. Similarly, an important application of NRQED is the calculation of the
parapositronium decay rate (see Refs. \cite{Kniehl:2000dh,Adkins:2003eh,Czarnecki:1999gv,Czarnecki:1999ci,Kniehl:2008ia,Kniehl:2009pg,Kniehl:2008dt}).

The analytic structure of the sunrise topology considered in this paper has been studied by means of differential equations in  \cite{Kotikov:1990kg,Kotikov:1991hm,Kotikov:1991pm,Bern:1993kr,Remiddi:1997ny}
and by using an effective-mass analysis in \cite{Kotikov:1990zs,Kniehl:2012hn,Kotikov:2020ccc}. Moreover, this integral family admits a closed-form solution in terms of ${}_3F_{2}$-hypergeometric functions as shown in \cite{Kalmykov:2008ge} (the corresponding off-shell diagrams with equal masses are considerably more complicated and their explicit solution requires Appell's $F_2$ hypergeometric functions \cite{Tarasov:2006nk}). In this paper we consider a finite representative of this integral family and we derive an expression in terms of eMPLs valid to all orders of the dimensional regulator. 

The  paper is organised as follows. In Section \ref{sec:sunrise},
we define the sunrise integral family and we present the ${}_3F_{2}$-hypergeometric representation for the finite sunrise integral considered in this paper. In Section \ref{sec:integral rep} we present a double integral representation for the relevant ${}_3F_{2}$-hypergeometric functions.
In Section \ref{sec:all orders empls}, we review properties of eMPLs and we present our all orders result for the sunrise integral. In Section \ref{sec:conclusions} we draw our conclusions. In Appendix \ref{App:intrep},
we present the detailed derivation of the double integral representations for the relevant ${}_3F_{2}$-hypergeometric functions. In Appendix \ref{App:pole cancellation} we provide a representation for the sunrise integral in terms of one-dimensional integrals over polylogarithmic expressions, at leading orders in the dimensional regulator. In Appendix \ref{app:definitions} we provide definitions relevant to the main results of this paper.

\section{The sunrise integral}
\label{sec:sunrise}
Following Ref. \cite{Kalmykov:2008ge} we
study the sunrise integral topology defined as,
\begin{equation}
    J_{i_1,i_2,i_3}(m^2,M^2)=\left.\int\int \frac{d^D k_1 d^D k_2}{[k_2^2-m^2]^{i_1}[k_1^2-M^2]^{i_2}[(k_1-k_2-q)^2-M^2]^{i_3}}\right|_{q^2=-m^2},
\end{equation}
with $D=4-2\epsilon$. This integral family has three master integrals, which can be chosen to be $J_{1,1,1},\,J_{1,1,2},\,J_{1,2,2}$. In this paper we consider the representative finite integral $J_{1,2,2}$ which can be solved in closed form in terms of hypergeometric functions \cite{Kalmykov:2008ge} as,
\begin{align}
  J_{1,2,2}(m^2,M^2)=&\hat{N}_{1}
  \frac{(1+\epsilon)}{\epsilon (1-\epsilon)}\times
  \left[ \frac{1}{6} \Hf{4}{3}{1,\frac{3}{2},1+\frac{\epsilon}{2},\frac{3}{2}+\frac{\epsilon}{2}}{2-\epsilon,\frac{5}{4},\frac{7}{4}}{-\frac{m^4}{4M^4}} \right.\nonumber\\
    &-\left(\frac{M^2}{m^2}\right)^{1-\epsilon}\frac{\epsilon}{(1+\epsilon)(1+2 \epsilon)}\Hf{4}{3}{1,\frac{1}{2}+\epsilon,1+\frac{\epsilon}{2},1+\epsilon}{\frac{3}{2}-\frac{\epsilon}{2},\frac{3}{4}+\frac{\epsilon}{2},\frac{5}{4}+\frac{\epsilon}{2}}{-\frac{m^4}{4M^4}}\nonumber\\
    &\left. -\left(\frac{M^2}{m^2}\right)^{-\epsilon}\frac{1-\epsilon}{(2-\epsilon)(3+2 \epsilon)}\Hf{4}{3}{1,\frac{3}{2}+\frac{\epsilon}{2},1+\epsilon,\frac{3}{2}+\epsilon}{2-\frac{\epsilon}{2},\frac{5}{4}+\frac{\epsilon}{2},\frac{7}{4}+\frac{\epsilon}{2}}{-\frac{m^4}{4M^4}}\right],
\label{J122}
\end{align}
where the normalization constant is,
\be
\hat{N}_{1} =\frac{\Gamma^2(1+\epsilon)(m^2)^{1-\epsilon}}{ (-M^2)^{-2-\epsilon}} \, .
\label{hN122}
\ee
Master integrals  $J_{1,1,1}$ and $J_{1,1,2}$ admit a similar representation in terms of hypergeomtric functions \cite{Kalmykov:2008ge} and will be studied elsewhere.

\section{Integral representations}
\label{sec:integral rep}
In this section we show that the hypergeometric functions of Eq.~(\ref{J122}) admit a two-fold integral representation. The derivation of this result relies on multiple identities for hypergeometric functions \cite{Kniehl:2005bc,Kniehl:2019vwr} and is presented in Appendix \ref{App:intrep}.  This representation will be used in the next sections to derive an explicit expression for the sunrise integral considered in this paper valid to all orders of the dimensional regulator in terms of eMPLs. 

By introducing the dimensionless ratio,

\be
t=\frac{m^2}{2 M^2}\,,
\label{t}
\ee
we find,
\be
J_{1,2,2} = \hat{N}_1 \biggl[J^{(1)}_{1,2,2}(t) - (2t)^{\ep-1} \, J^{(2)}_{1,2,2}(t) - (2t)^{\ep} \, J^{(3)}_{1,2,2}(t) \biggr] \, ,
\label{J122N}
\ee
where,
\bea
&&J^{(1)}_{1,2,2}(t) = \frac{1+\ep}{6\ep (1-\ep)} \, \Hf{4}{3}{1,\frac{3}{2},1+\frac{\epsilon}{2},\frac{3}{2}+\frac{\epsilon}{2}}{2-\epsilon,\frac{5}{4},\frac{7}{4}}{-t^2} = - \frac{\hat{K}}{2^{2\epsilon+2}\ep t^2} \, I^{(1)}(t)\,, \nonumber \\
&&J^{(2)}_{1,2,2}(t) = \frac{1}{(1+2\ep)(1-\ep)}\,  \Hf{4}{3}{1,\frac{1}{2}+\epsilon,1+\frac{\epsilon}{2},1+\epsilon}{\frac{3}{2}-\frac{\epsilon}{2},\frac{3}{4}+\frac{\epsilon}{2},\frac{5}{4}+\frac{\epsilon}{2}}{-t^2} =  \frac{\hat{K}}{2^{4\epsilon+1} t^{1-\ep}} \, I^{(2)}(t)\, , \nonumber \\
&&J^{(3)}_{1,2,2}(t) = \frac{1+\ep}{\ep (2-\ep)(3+2\ep)} \,\Hf{4}{3}{1,\frac{3}{2}+\frac{\epsilon}{2},1+\epsilon,\frac{3}{2}+\epsilon}{2-\frac{\epsilon}{2},\frac{5}{4}+\frac{\epsilon}{2},\frac{7}{4}+\frac{\epsilon}{2}}{-t^2} = - \frac{\hat{K}}{2^{4\epsilon+2}\ep t^2} \, I^{(3)}(t)\,,
\label{Ji122N1}
\eea
where $\hat{K}$ is defined as,
\begin{equation}
\hat{K}=\frac{\Gamma(1-\ep)}{\Gamma(1-2\ep)\Gamma(1+\ep)},
\end{equation}
while the factors $I^{(i)}(t)$ represent the relevant double integrals,
\be
\label{I1,2}
I^{(1)}(t)=I_1^{(1)}(t)-\epsilon I_2^{(1)}(t),\; I^{(2)}(t)=I_1^{(2)},\; I^{(3)}(t)= I_1^{(3)}(t)-\frac{\epsilon}{2}I_2^{(3)}(t)\,,
\ee
with,
\begin{align}
\label{eq:2integrals p}
     I^{(1)}_1(t)&=\int_0^1 \,dp \,p^{\epsilon -1}(1-p)^{-\epsilon -\frac{1}{2}} \left((p^2 t^2+1)^{-\frac{1}{2}}-1\right)  ,\nonumber\\
    I^{(1)}_2(t)&= \int_0^1\, dp\,p^{\epsilon -1}\,(1-p)^{-\epsilon -\frac{1}{2}}\,(p^2 t^2+1)^{-\frac{1}{2}}  \, q(p)^{\epsilon }\int_0^{q(p)} dz\left((1-z)^{-\frac{1}{2}}-1\right) z^{-\epsilon -1},\nonumber\\
    I^{(2)}_1(t)&=  \int_0^1 \,dp\, p^{3 \epsilon -1} (1-p)^{-\epsilon -\frac{1}{2}}  \left(p^2 t^2+1\right)^{-\epsilon -\frac{1}{2}} \int_{0}^{pt} dz \,z^{-\epsilon } \left(z^2+1\right)^{\epsilon -\frac{1}{2}},\nonumber\\
     I^{(3)}_1(t)&=\int_0^1\, dp \,p^{2 \epsilon -1}(1-p)^{-\epsilon -\frac{1}{2}} \left(\left(p^2 t^2+1\right)^{-\frac{\epsilon}{2} -\frac{1}{2}}-1\right),\nonumber\\
    I^{(3)}_2(t)&=\int_0^1 \,dp \,p^{2 \epsilon -1} (1-p)^{-\epsilon -\frac{1}{2}} \left(p^2 t^2+1\right)^{-\frac{\epsilon}{2} -\frac{1}{2}} \,q(p)^{\frac{\epsilon}{2}}\int_0^{q(p)} dz \left((1-z)^{-\frac{\epsilon}{2} -\frac{1}{2}}-1\right) z^{-\frac{\epsilon }{2}-1},
\end{align}
%
and the upper integration bound is,

\be
q(p)=\frac{p^2 t^2}{p^2 t^2+1}.
\label{y}
\ee

We remark that integral $J_{1,2,2}$ has a finite $\epsilon$ expansion. While the representation above does not make this fact manifest, we obtain a manifestly finite expression in Appendix \ref{App:pole cancellation}.


\section{All orders result in terms of elliptic polylogarithms}
\label{sec:all orders empls}

In this section we derive an eMPLs representation for the sunrise integral $J_{1,2,2}$ valid to all orders of the dimensional regulator. Specifically, we start with a short review of eMPLs, discussing their definition and the basic analytic properties. We then discuss the general structure of the integral representations presented in the previous section and we show that, by defining a new integration variable, their dependence on the relevant elliptic curve can be made explicit. We conclude by discussing the general solution strategy used to express these integrals in terms of eMPLs to all orders of the dimensional regulator, and present our final results.

\subsection{Elliptic polylogarithms}
\label{sec: TH Framework}
We are interested in the computation of iterated integrals of the form,
\begin{equation}
\label{eq:GenericItIntEll}
    \int_0^{x} dx_1 R_1(x_1,y(x_1)) \int_0^{x_1} dx_2 R_2(x_2,y(x_2))\dots\int_0^{x_{n-1}} dx_n R_{n}(x_n,y(x_n))\, ,
\end{equation}
where $R_i$ are rational functions of their arguments and $y(x)$ is an elliptic curve,
\begin{equation}
y(x)=\sqrt{(x-a_1)(x-a_2)(x-a_3)(x-a_4)}\,,
\end{equation}
All iterated integrals of the form (\ref{eq:GenericItIntEll}) can be expressed in terms of eMPLs.
In the complex plane, eMPLs are defined as
\begin{equation}
\label{eq:E4_def}
\Ef{n_1,  \dots,  n_k}{c_1,  \dots, c_k}{x} = \int_0^xdt\,\varphi_{n_1}(c_1,t)\,\Ef{n_2,  \dots , n_k}{c_2,  \dots, c_k}{t}\, ,
\end{equation}
with $n_i\in\mathbb{Z}$ and $c_i\in\mathbb{C}\cup \{\infty\}$. The recursion starts at $\Ef{}{}{x}=1$. By construction, the kernels $\varphi_n(c,x)$ have at most simple poles, and they are (see \cite{Broedel:2017kkb} for a detailed discussion), 
\begin{align}
\label{PhiFunc}
& \varphi_0(0,x)= \frac{c_4}{y(x)}\, ,  \nonumber\\&
\varphi_1(c,x)= \frac{1}{x-c}\, , \qquad
\varphi_{-1}(c,x) = \frac{y(c)}{(x-c)y(x)}-(\delta_{c0}+\delta_{c1})\frac{1}{x-c}\, ,  \nonumber\\& \varphi_{-1}(\infty,x) = \frac{x}{y(x)}\, , \qquad
\varphi_1(\infty,x) = \frac{c_4}{y(x)}\,Z_4(x)\, , \nonumber\\&
\varphi_n(\infty,x)= \frac{c_4}{y(x)}\,Z_4^{(n)}(x)\, , \nonumber\\&
\varphi_{-n}(\infty,x) = \frac{x}{y(x)}\,Z_4^{(n-1)}(x)-\frac{\delta_{n2}}{c_4}\, , \nonumber\\&
\varphi_n(c,x) = \frac{1}{x-c}\,Z_4^{(n-1)}(x)-\delta_{n2}\,\Phi_4(x)-(\delta_{c0}+\delta_{c1})\frac{Z_4^{(n-1)}(c)}{x-c}\, ,
\nonumber\\&
\varphi_{-n}(c,x) = \frac{y(c)}{(x-c)y(x)}\,Z_4^{(n-1)}(x)-(\delta_{c0}+\delta_{c1})\frac{Z_4^{(n-1)}(c)}{x-c}\, ,\quad (n>1)
\end{align}
where $y(c)$ and $c_4$ are independent of $x$ with,
\begin{equation}
 c_{4} = \frac{1}{2}\sqrt{a_{13}a_{24}} \quad \text{with} \quad a_{ij}=a_i-a_j\, .
\end{equation}   
Moreover,
\begin{equation}
    \Ef{\vec{1}}{\vec{0}}{x}\equiv\frac{\log(x)^n}{n!},
\end{equation}
where $\vec{1}$ and $\vec{0}$ are vectors with entries equal to $1$ and $0$ respectively, and $n=\text{length}(\vec{1})=\text{length}(\vec{0})$. The function $Z_4(x)$ is defined by first introducing an auxiliary function $ \Phi_4(x)$, 
\begin{equation}
\label{eq:Phi4}
    \Phi_4(x)\equiv \widetilde{\Phi}_4(x)+4c_4\frac{\eta_1}{\omega_1}\frac{1}{y} = \frac{1}{c_4\,y} \left( x^2 - \frac{s_1}{2}\,x + \frac{s_2}{6} \right)+4c_4\frac{\eta_1}{\omega_1}\frac{1}{y}\, ,
\end{equation}
where $\eta_1, \omega_1, s_1, s_2, c_4$ are independent of $x$ and they are defined in \cite{Broedel:2017kkb}, and,
\begin{equation}
  \widetilde{\Phi}_4(x)= \frac{1}{c_4 y}\left(x^2-\frac{s_1 x}{2}+\frac{s_2}{6}\right)\, , 
\end{equation}
whose primitive is,
\begin{equation}
\label{eq:Z4=intPhi4}
  Z_4(x)= \int_{a_1}^x dt \, \Phi_4(t)\, . 
\end{equation}

Elliptic polylogarithms are a generalisation of ordinary multiple polylogarithms (MPLs), defined recursively as,
\begin{equation}
    G(a_1,a_2,\ldots,a_n;x)=\int_0^x \frac{dt}{t-a_1} G(a_2,\ldots,a_n,t),
\end{equation}
with $G(;x)\equiv 1$ and,
\begin{equation}
    G(\vec{0},x)\equiv\frac{\log(x)^n}{n!}\,.
\end{equation}
By definition we see that MPLs are a subset of eMPLs,
\begin{equation}
    \Ef{1,  \dots,  1}{c_1,  \dots, c_n}{x} = G(c_1,c_2,\dots,c_n;x)\, ,
\end{equation}
where $c_i \neq \infty$.

The appearance of the function $Z_4(x)$ and its regularised powers $Z_4(x)^{(n)}$ is due to the requirement that, in analogy to MPLs, integration kernels have at most simple poles \cite{Broedel:2017kkb}. The integrals considered in this paper do not involve $Z_4(x)$ and its regularised powers.

As for all iterated integrals, eMPLs satisfy a shuffle algebra, with the shuffle product defined as,
\begin{equation}
 \Efff{a_1}{a'_1}{\dots}{\dots}{a_n}{a'_n}{x}\Efff{b_1}{b'_1}{\dots}{\dots}{b_m}{b'_m}{x} =  \sum_{\vec{c}=\vec{a}\shuffle\vec{b}} \Efff{c_1}{c'_1}{\dots}{\dots}{c_{n+m}}{c'_{n+m}}{x}\, .
\end{equation}
The vector $\vec{c}$ is the vector obtained by performing all the shuffles of $\vec{a}$ and $\vec{b}$, preserving the ordering of the elements of $\vec{a}$ and $\vec{b}$ respectively.

\subsection{Regularisation}
\label{sec:regularisation}
As we will see in the next sections we are interested in computing definite integrals of the form,
\begin{equation}
\label{eq:reg_ex}
    \int_0^1 f(x)dx= F(1)-F(0) , \quad \frac{\partial F(x)}{\partial x}=f(x)\,.
\end{equation}
In some cases the primitive is ill defined when computed at the integration bounds, and in order to compute the definite integral one needs to perform two limits,
\begin{equation}
\label{eq:reg_exlim_0}
    \int_0^1 f(x)dx= \lim_{x\rightarrow 1}F(x)-\lim_{x\rightarrow 0}F(x).
\end{equation}
We illustrate how the limits are performed for the case,
\begin{equation}
\label{eq:prim_ex}
   F(x)= \Ef{-1}{-1}{1} \Ef{1}{1}{x}+\Eff{1}{0}{-1}{-1}{x}-\Eff{1}{1}{-1}{-1}{x}.
\end{equation}
 From Eq.(\ref{eq:prim_ex}) we see that $F(0)=0$, while two eMPLs, $\Ef{1}{1}{x}$ and $\Eff{1}{1}{-1}{-1}{x}$, are divergent for $x=1$, therefore the computation of the integral (\ref{eq:reg_ex}) requires the computation of the following limit,
\begin{equation}
\label{eq:reg_exlim}
    \int_0^1 f(x)dx= \lim_{x\rightarrow 1}F(x).
\end{equation}
The limits can be computed by performing shuffle regularisation. Specifically, we first use shuffle identities to isolate the logarithmic divergences,
\begin{equation}
    \Eff{1}{1}{-1}{-1}{x}=\Ef{-1}{-1}{x} \Ef{1}{1}{x}-\Eff{-1}{-1}{1}{1}{x}\,,
\end{equation}
and by inserting this result in (\ref{eq:prim_ex}) we find,
\begin{equation}
     F(x)=\Ef{-1}{-1}{1} \Ef{1}{1}{x}-\Ef{-1}{-1}{x} \Ef{1}{1}{x}+\Eff{-1}{-1}{1}{1}{x}+\Eff{1}{0}{-1}{-1}{x}.
\end{equation}
The divergences in $x=1$ are now only carried by factors of $\Ef{1}{1}{x}=\log(1-x)$. However it is easy to see that the divergences explicitly cancel in the limit, and we obtain,
\begin{equation}
    \int_0^1 f(x)dx= \lim_{x\rightarrow 1}F(x) =\Eff{-1}{-1}{1}{1}{1}+\Eff{1}{0}{-1}{-1}{1}.
\end{equation}

We remark that for the choice of the integration kernels (\ref{PhiFunc}), divergent eMPLs at the integration bounds considered in this work, i.e. $0$ and $1$,  are those of the form,
\begin{equation}
    \Eff{1}{1}{\vec{m}}{\vec{n}}{1}, \quad \Efff{1}{0}{\dots}{\dots}{1}{0}{0},\quad \col{\vec{m}}{\vec{n}}\neq \col{1,\dots ,1}{0,\dots0},
\end{equation}
implying that only $\Eff{1}{1}{\vec{m}}{\vec{n}}{1}$ need to be shuffle regulated. As discussed above, after shuffle regularisation, the diverging terms are powers of $\Ef{1}{0}{0}$ and $\Ef{1}{1}{1}$, and a manifestly finite result is obtained by taking the limit $x\rightarrow 1$ and $x\rightarrow 0$. For a finite result all divergent factors are multiplied by a vanishing expression in the limits.

In what follows we use a more direct and equivalent procedure to compute definite integrals involving singular primitives. Specifically, we formally compute definite integrals as the difference of the primitive evaluated at the integration bounds. Finally, we shuffle regulate the result and set to zero all divergent logarithms.

\subsection{Elliptic polylogarithms and all orders result}
In this section we show that the double integrals of Section \ref{sec:sunrise} can be expressed as a $\mathbb{Q}$-linear combinations of eMPLs to all orders of the dimensional regulator. We start by observing that the integrals of (\ref{eq:2integrals p}) have the following general form\footnote{Integrals $I_1^{(1)}$ and $I_1^{(3)}$ are in fact one-fold integrals, but for the general discussion of this section includes trivially the simpler case},
\begin{equation}
\label{eq: gen double int}
    \int_0^1 d p \sum_{i=1}^{n_1} f_{1,i}(p)g_{1,i}(p)^\epsilon \int_0^{h(p)} \sum_{j=1}^{n_2} f_{2,j}(z)g_{2,j}(z)^\epsilon dz, \quad n_1,n_2 \in \mathbb{N},
\end{equation}
where $f_{1,i}(p),g_{1,j}(p)$ are algebraic functions of $p$, with algebraic factors,
\begin{equation}
\sqrt{1-p},\;\sqrt{1+t^2 p^2}, \quad  0<p<1,\; 0<t<1,
\end{equation}
while $h(p)$ is a rational function of $p$ (see below for the analytic properties of $f_{2,i}(p),g_{2,j}(p)$). The dependence on the relevant elliptic curve can be made explicit by the variable change,
\begin{equation}
\label{eq:cov}
    p(x)=1-x^2,\quad 0<x<1,
\end{equation}
so that,
\begin{equation}
    -\int_0^1 d x \sum_{i=1}^{n_1} f'_{1,i}(x,y(x))g'_{1,i}(x,y(x))^\epsilon \int_0^{h(p(x))} \sum_{j=1}^{n_2} f_{2,j}(z)g_{2,j}(z)^\epsilon dz\,,
\end{equation}
where the elliptic curve is,
\begin{equation}
    y(x)=\sqrt{\frac{1}{t^2} + (1 - x^2)^2},
\end{equation}
and,
\begin{equation}
    f'_{1,i}(x,y(x))= \frac{d p(x)}{d x}f_{1,i}(p(x)),\quad g'_{1,i}(x,y(x))= g_{1,i}(p(x))\,.
\end{equation}
For the integrals considered in this work, $f'_{1,i}(x,y(x))$ is a rational function on the elliptic curve $y(x)$ with at most simple poles, implying that it can be expressed as a $\mathbb{Q}$-linear combination of kernels $\varphi_{0}(0,x)$, $\varphi_{1}(c,x)$, $\varphi_{-1}(c,x)$, $\varphi_{-1}(\infty,x)$. On the other hand $g'_{1,i}(x,y(x))$ is a rational function on the elliptic curve of the form,
\begin{equation}
\label{eq:logarg}
    g'_{1,i}(x,y(x))=y(x)^\alpha R_{1,i}(x),\quad \alpha\in \mathbb{N},
\end{equation}
where $R_{1,i}$ is a rational function, which implies that upon expanding in powers of $\epsilon$, the resulting logarithmic factor can be evaluated in terms of eMPLs of the form $\Ef{1}{c}{x}$, i.e. ordinary MPLs. The $\epsilon$ expansion is performed according to,
\begin{equation}
\label{eq: eps log exp}
   g'_{1,i}(x,y(x))^\epsilon=\sum_{i=0}^\infty \frac{\log (g'_{1,i}(x,y(x)))^i \epsilon^i}{i!}.
\end{equation}
The logarithm can be expressed as $\mathbb{Q}$-linear combinations of eMPLs, by using the general identity,
\begin{equation}
\label{eq: log2empls}
    \log(a(x,y(x)))=\log(a(0,y(0)))+\int_0^x dz \frac{d a(z,y(z))}{d z}\frac{1}{a(z,y(z))},
\end{equation}
for regular and non-vanishing $a(0,y(0))$. As anticipated, eq.~(\ref{eq:logarg}) implies that the integrand on the right hand side can be expressed as a $\mathbb{Q}$-linear combination of $\varphi_{1}(c,x)$. For a subset of integrals we have that $a(x,y(x))=x^\beta r(x,y(x))$, with regular and non-vanishing $r(0,y(0))$, $\beta\neq 0$, and the identity above is ill-defined. Nonetheless in these cases we simply have,
\begin{equation}
    \log(a(x,y(x)))=\log(r(0,y(0)))+\beta\log(x)+\int_0^x dz \frac{d r(z,y(z))}{d z}\frac{1}{r(z,y(z))}.
\end{equation}

Results of the same form can be obtained for the inner integral. This is achieved by changing the upper integration bound to $x$, by using the fact that,
\begin{equation}
    \int_0^{a(x)}b(z) dz=\int_0^{a(1)}b(z)+\int_1^x dz \frac{\partial}{\partial x}\int_0^{a(x)}b(z) dz,
\end{equation}
and
\begin{equation}
\frac{\partial}{\partial x}\int_0^{a(x)}b(z) dz =b(a(z)) \frac{d b(z)}{d z},
\end{equation}
where the choice of unit lower integration bound is conventional at this point, and will be clarified later. For the case at hand, the identities above imply,
\begin{align}
\label{eq:der int inner}
\int_0^{h(p(x))} \sum_{j=1}^{n_2} f_{2,j}(z)g_{2,j}(z)^\epsilon dz&=\int_1^x \sum_{j,k=1}^{n_2}f_{2,j}(h(p(z)))g_{2,j}(h(p(z)))^\epsilon \frac{d(f_{2,k}(z)g_{2,k}(z)^\epsilon)}{d z} dz\nonumber\\
&+\int_0^{h(p(1))}\sum_{j=1}^{n_2}f_{2,j}(z)g_{2,j}(z)^\epsilon dz,
\end{align}
For the $h(p(x))$ considered in this work, $h(p(1))=0$, so that the last integral on the right hand side vanishes. Moreover, for the relevant $f_{2,j}(z)$ and $g_{2,j}(z)$, we can write,
\begin{equation}
    \int_1^x \sum_{j,k=1}^{n_2}f_{2,j}(h(p(z)))g_{2,j}(h(p(z)))^\epsilon \frac{d(f_{2,k}(z)g_{2,k}(z)^\epsilon)}{d z} dz=\int_1^x \sum_{i=1}^n f'_{2,i}(z,y(z))g'_{2,i}(z,y(z))^{\epsilon} dz\,,
\end{equation}
where $f'_{2,i}(z,y(z))$ are rational functions on $y(z)$ with at most simple poles while  $g'_{2,i}(z,y(z))$ have the same form as eq.~(\ref{eq:logarg}). Logarithms and prefactors may be expressed as $\mathbb{Q}$-linear combinations of eMPLs and integration kernels respectively as discussed in the previous paragraph. 

This analysis implies that integrals of the form eq.~(\ref{eq: gen double int}) formally evaluate to,
\begin{equation}
\label{eq: gen eMPLs 2-integral}
   \sum_{l=1}^{n'} C_l\sum_{i,j=0}^\infty \frac{\epsilon^{i+j}}{i!j!}\int_0^1 dx k_{1,l}(x) L_{1,l}(x)^i\int_1^x dz k_{2,l}(z) L_{2,l}(z)^j,\quad n'\in \mathbb{N},\; C_l\in \mathbb{Q}\,,
\end{equation}
where $L_{i,j}(x)$ are $\mathbb{Q}$-linear combinations of eMPLs of depth one, while $k_{i,j}(x)$ are $\mathbb{Q}$-linear combinations of integration kernels. These integrals can be directly evaluated in terms of eMPLs by shuffle expanding products of eMPLs of the integrands, and by using recursively the definition of eMPLs. 

Eq.~(\ref{eq: gen eMPLs 2-integral}) is not well defined in general, since double integrals might be individually divergent, while only the full linear combination corresponds to a finite result. Nonetheless a manifestly finite result can be obtained by performing shuffle regularisation as described in \ref{sec:regularisation}. More precisely, diverging powers of logarithms (in our case powers of $\Ef{1}{0}{0}$ and $\Ef{1}{1}{1}$) are isolated by using shuffle identities and are subsequently set to zero.

In order to make the notation more compact, and make properties of the result in terms of eMPLs manifest, we use the following notation for the double integrals of (\ref{eq: gen eMPLs 2-integral}). By denoting a primitive of $k_{i,j}(x)$ as $K_{i,j}(x)$, and by defining the $*$-operator as,
\begin{equation}
\Ef{\vec{n}}{\vec{c}}{x}*  \Ef{\vec{m}}{\vec{d}}{x}=    \Eff{\vec{n}}{\vec{c}}{\vec{m}}{\vec{d}}{x},
\label{star}
\end{equation}
we can write
\begin{equation}
\label{eq: gen eMPLs * notation}
   \Reglim_{0,1} \sum_{l=1}^{n'} C_l\sum_{i,j=0}^\infty \frac{\epsilon^{i+j}}{i!j!} K_{1,l}(x) * L_{1,l}(x)^i \left[ K_{2,l}(z) * L_{2,l}(z)^j\right]_1^x,
\end{equation}
where all products of eMPLs are shuffle expanded before applying the $*$-operator, and these operations are carried out for the inner square brackets first. Finally, the lower and upper scripts applied to the square brackets denote the following operation,
\begin{equation}
    \left[F(x)\right]_1^x= F(x)-F(1)\,,
\end{equation}
while the regularisation operator is,
\begin{equation}
    \Reglim{}_{0,1} F(x)=\Reglim_{x\rightarrow 1}F(x)-\Reglim_{x\rightarrow 0}F(x)\,.
\end{equation}

By applying the procedure described above we obtain one of the main results of this paper, i.e. an explicit expression for the integrals of Eq.~(\ref{eq:2integrals p}) in terms of eMPLs valid to all orders of the dimensional regulator,
\begin{align}
    I_1^{(1)}(t)&= \Reglim_{0,1}\sum_{i=0}^\infty \frac{\epsilon ^{i}}{i!} K_1 * L_1^{i}\,,\nonumber\\
    I_2^{(1)}(t)&=\Reglim_{0,1}\sum_{i,j=0}^\infty\frac{\epsilon ^{i+j}}{i!j!}K_2 * L_3^{i}  \left[K_3 * L_2^{j}\right]_1^x\,,\nonumber\\
    I_1^{(2)}(t)&=\Reglim_{0,1}\sum_{i,j=0}^\infty\frac{\epsilon ^{i+j}}{i!j!}K_4*  L_5^{i}  \left[K_5 * L_4^{j}\right]_1^x\,,\nonumber\\
   I_1^{(3)}(t)&=\Reglim_{0,1}\sum_{i=0}^\infty\frac{\epsilon ^{i}}{i!}K_6 *L_6^{i} +\Reglim_{0,1} \sum_{i=0}^\infty\frac{\epsilon ^{i}}{i!}K_4 * L_7^{i}\,,\nonumber\\
   I_2^{(3)}(t)&=\Reglim_{0,1}\sum_{i,j=0}^\infty\frac{\epsilon ^{i+j}}{i!j!}K_7*  L_8^{i}  \left[K_9 * L_4^{j}\right]_1^x+\Reglim_{0,1}\sum_{i,j=0}^\infty\frac{\epsilon ^{i+j}}{i!j!} K_7* L_8^{i}  \left[K_8 * L_9^{j}\right]_1^x \, ,
\label{main}
\end{align}
where $K_i$ and $L_i$ are depth one eMPLs and their definition is provided in Appendix C.

\subsection{Example}

We show how the solution strategy of the previous section works in practice by considering integral $I^{(2)}_1(t)$. The dependence on the elliptic curve is made explicit by applying the variable change (\ref{eq:cov}), 
\begin{equation}
\label{eq:I2}
I^{(2)}_1(t)=\int_0^1 dx \;\frac{2}{t \left(1-x^2\right) y(x)}\left(\frac{\left(1-x^2\right)^3}{t^2 x^2 y(x)^2}\right)^{\epsilon } 
   \int_0^{t(1- x^2)}dz \frac{1}{\sqrt{z^2+1}} \left(z+\frac{1}{z}\right)^{\epsilon },
\end{equation}
By applying eq.~(\ref{eq:der int inner})  the inner integral can be expressed as
\begin{equation}
   \int_0^{t-t x^2}dz \frac{1}{\sqrt{z^2+1}} \left(z+\frac{1}{z}\right)^{\epsilon }=-\int_1^x dz \frac{2 z}{y(z)} \left(\frac{t y^2(z)}{1- z^2}\right)^{\epsilon } .
\end{equation}
All the $\epsilon$-powers can be expanded in $\epsilon$ by eq.~(\ref{eq: eps log exp}). For example we have 
\begin{equation}
\left(\frac{t y^2(x)}{1- x^2}\right)^{\epsilon }=\sum_{i=0}^\infty\frac{\epsilon ^i }{i!}\log ^i\left(\frac{t y^2(x)}{1- x^2}\right)\, ,
\end{equation}
The resulting logarithm can be expressed in terms of eMPLs by eq.~(\ref{eq: log2empls}), 
\begin{equation}
  \log\left(\frac{t y^2(x)}{1- x^2}\right)= \log \left(t^2+1\right)-\log (t)+ \int_0^x dz \frac{2 z \left(t^2 \left(z^2-1\right)^2-1\right)}{t^2 \left(z^2-1\right) y(z)^2}\,.
\end{equation}
The integrand above can be written in terms of the integration kernels as,
\begin{equation}
    \frac{2 z \left(t^2 \left(z^2-1\right)^2-1\right)}{t^2 \left(z^2-1\right) y(z)^2}=\sum_{i=1}^4\varphi _1\left(a_i,z\right)-\varphi _1(-1,z)-\varphi _1(1,z)\,,
\end{equation}
where we denoted with $a_i$ the four roots of the elliptic curve,
\begin{equation}
    a_1=-\frac{\sqrt{t-i}}{\sqrt{t}}\,,\; a_2=\frac{\sqrt{t-i}}{\sqrt{t}}\,, \; a_3=-\frac{\sqrt{t+i}}{\sqrt{t}}\,, \; a_4=\frac{\sqrt{t+i}}{\sqrt{t}}\,.
\end{equation}
Upon integration we find,
\begin{equation}
  \log\left(\frac{t y^2(x)}{1- x^2}\right)=   \sum_{i=1}^4\Ef{1}{a_i}{x}-\Ef{1}{-1}{x}-\Ef{1}{1}{x}+\log \left(t^2+1\right)-\log (t)\,.
\end{equation}
Finally, all prefactors can be expressed in terms of integration kernels, for example, referring to eq.~(\ref{eq:I2}),
\begin{equation}
    \frac{2}{t \left(1-x^2\right) y(x)}=\varphi _{-1}(-1,x)-\varphi _{-1}(1,x)-\varphi _1(1,x)\,.
\end{equation}
By applying these methods to all relevant logarithms and prefactors, we obtain a result in terms of integrals of the form (\ref{eq: gen eMPLs 2-integral}), which are directly evaluated to eMPLs by, e.g, Eq.~(\ref{eq: gen eMPLs * notation}),
\begin{equation}
   I_1^{(2)}(t)=\Reglim_{0,1}\sum_{i,j=0}^\infty\frac{\epsilon ^{i+j}}{i!j!}K_4*  L_5^{i}  \left[K_5 * L_4^{j}\right]_1^x\,,
\end{equation}
where the definitions of $L_i$ and $K_i$ are provided in Appendix \ref{app:definitions}.

\section{Conclusions}
\label{sec:conclusions}
In this paper we studied a sunrise integral with two different internal masses and pseudo-threshold kinematics in dimensional regularisation. This integral admits a closed-form solution in terms of hypergeometric functions \cite{Kalmykov:2008ge} and we use this representation as the starting point of our analysis. Specifically, we show that all relevant hypergeometric functions admit a representation in terms of double iterated integrals depending on one elliptic curve and no other algebraic functions. When expanding these integrals with respect to the (vanishing) dimensional regulator, the coefficients of the expansion are iterated integrals over rational functions on the relevant elliptic curve, with at most simple poles. We derive an expression for the sunrise integral valid to all orders of the dimensional regulator in terms of eMPLs. 

A similar analysis can be carried out for all the master integrals of the sunrise topology considered in this paper. However the explicit solution in terms of elliptic polylogarithms to all orders of the dimensional regulator is complicated by the appearance of integration kernels with higher poles, and we leave this study for future work.

\newpage

\appendix
\def\theequation{A\arabic{equation}}
\setcounter{equation}{0}

\section{Integral representations for the hypergeometric functions}
\label{App:intrep}

In this Section we provide the detailed derivation of the integral representations for the ${}_4F_3$-hypergeometric functions of Eq.~(\ref{J122}), identities Eq.~(\ref{Ji122N1}). Similar techniques have been used in Refs. \cite{Kniehl:2005bc,Kniehl:2019vwr,Fleischer:1997bw,Fleischer:1998nb,Fleischer:1999hp,Kotikov:2007vr,Davydychev:2003mv}. In what follows we use the following definitions and identities,
  \begin{equation}
\hat{K}=\frac{\Gamma(1-\ep)}{\Gamma(1-2\ep)\Gamma(1+\ep)},\;\;
\hat{K}_1=\frac{\Gamma \left(\frac{1}{2}\right)}{\Gamma \left(\frac{1}{2}-\epsilon \right) \Gamma (\epsilon +1)}= \frac{\hat{K}}{2^{2\ep}},
\;\;\hat{K}_2=\frac{\Gamma \left(\epsilon +\frac{1}{2}\right)}{\Gamma \left(\frac{1}{2}-\epsilon \right) \Gamma (2 \epsilon +1)}
= \frac{\hat{K}_1}{2^{2\ep}}= \frac{\hat{K}}{2^{4\ep}},
\label{Ki}
\end{equation}
and, referring to Eq.~(\ref{eq:2integrals p}), we define the integrals $J_i(p)$ $(i=1,2,3)$ as,
\begin{align}
\label{Ji}
    J_1(p)&=q(p)^{\epsilon }\int_0^{q(p)} dz\left((1-z)^{-\frac{1}{2}}-1\right) z^{-\epsilon -1},\;\; J_2(p)=\int_{0}^{pt} dz \,z^{-\epsilon } \left(z^2+1\right)^{\epsilon -\frac{1}{2}},\nonumber\\
    J_3(p)&=q(p)^{\frac{\epsilon}{2}}\int_0^{q(p)} dz \left((1-z)^{-\frac{\epsilon}{2} -\frac{1}{2}}-1\right) z^{-\frac{\epsilon }{2}-1}.
\end{align}

\subsection*{The first hypergeometric function}

The first ${}_4F_3$-hypergeometric function of $J_{1,2,2}$ admits the following series representation,
\bea
&&F_{1}(t) \equiv
{}_{4}F_{3}\left(\left.
\begin{array}{c}
1,\frac{3}{2},1+\frac{\ep}{2},\frac{3}{2}+\frac{\ep}{2}\\
2-\ep,\frac{5}{4},\frac{7}{4}
\end{array}
\right| -t^2 \right), \,
\nonumber \\
&&= \sum_{m=0}^{\infty} \, \frac{\Gamma(m+\frac{3}{2})\Gamma(m+1+\frac{\ep}{2})\Gamma(m+\frac{3}{2}+\frac{\ep}{2})}{\Gamma(m+2-\ep)
\Gamma(m+\frac{5}{4})\Gamma(m+\frac{7}{4})} \, \frac{\Gamma(2-\ep)
\Gamma(\frac{5}{4})\Gamma(\frac{7}{4})}{\Gamma(\frac{3}{2})\Gamma(1+\frac{\ep}{2})\Gamma(\frac{3}{2}+\frac{\ep}{2})}
\, {(-t^2)}^m \,,
\label{J122.1.1}
\eea
where $t=m^2/(2M^2)$ as defined in Eq.~(\ref{t}) of the main text. The product $\Gamma(\alpha)\Gamma(1/2+\alpha)$ can be written as,
\be
\Gamma(\alpha)\Gamma(1/2+\alpha) \,= \, 2^{1-2\alpha} \, \sqrt{\pi} \, \Gamma(2\alpha) \, ,
\label{Gamma}
\ee
which results in the following simplified expression for $F_{1}(t)$,
\be
\sum_{m=0}^{\infty} \, \frac{\Gamma(m+\frac{3}{2})\Gamma(2m+2+\ep)}{\Gamma(m+2-\ep)\Gamma(2m+\frac{5}{2})} \, \frac{\Gamma(2-\ep)
\Gamma(\frac{5}{2})}{\Gamma(\frac{3}{2})\Gamma(2+\ep)}
\, {(-t^2)}^m  \, .
\label{J122.1.2}
\ee
It is convenient to use the following integral representations for the ratio of gamma functions, 
\be
\frac{\Gamma(2m+2+\ep)}{\Gamma(2m+\frac{5}{2})} = \int_0^1 \, dp \, \frac{p^{2m+1+\ep}(1-p)^{-1/2-\ep}}{\Gamma(\frac{1}{2}-\ep)} \, .
\label{Int.p.1}
\ee
We find,
\be
F_{1}(t) = \int_0^1 \, dp \, \frac{p^{1+\ep}(1-p)^{-1/2-\ep}}{\Gamma(\frac{1}{2}-\ep)}
\sum_{m=0}^{\infty} \, \frac{\Gamma(m+\frac{3}{2})}{\Gamma(m+2-\ep)} \, \frac{\Gamma(2-\ep)
\Gamma(\frac{5}{2})}{\Gamma(\frac{3}{2})\Gamma(2+\ep)}
\, {(-(tp)^2)}^m  \, .
\label{J122.1.3}
\ee

In order to proceed with our analysis it is convenient to consider first the series on the right hand side in the limit $\ep=0$,
\be
\sum_{m=0}^{\infty} \, \frac{\Gamma(m+\frac{3}{2})}{(m+1)!}  \, {(-(tp)^2)}^m =
\sum_{m=1}^{\infty} \, \frac{\Gamma(m+\frac{1}{2})}{m!}  \, {(-(tp)^2)}^{m-1}
= - \frac{\Gamma(\frac{1}{2})}{(tp)^2} \left[ \frac{1}{\left(1+t^2p^2\right)^{1/2}} -1 \right]
\label{Sum.1.1}.
\ee
In the general case we have,
\bea
&&\sum_{m=0}^{\infty} \, \frac{\Gamma(m+\frac{3}{2})}{\Gamma(m+2-\ep)}  \, {(-(tp)^2)}^m  =
\sum_{m=1}^{\infty} \, \frac{\Gamma(m+\frac{1}{2})}{\Gamma(m+1-\ep)} \, {(-(tp)^2)}^{m-1} \nonumber \\
&&= - \frac{\Gamma(\frac{1}{2})}{\Gamma(1-\ep)(tp)^2} \left[ {}_{2}F_{1}\left(1, \frac{1}{2}; 1-\ep; -p^2t^2\right)
 \, -1 \right]
\label{Sum.1.2}.
\eea
Using standard properties of the ${}_{2}F_{1}$-function,
\be
{}_{2}F_{1}\Big(a,b;c;z\Bigr)= (1-z)^b  \,{}_{2}F_{1}\left(c-a,b;c;\frac{z-1}{z}\right) \,
\label{2F1.1},
\ee
we obtain,
\be
{}_{2}F_{1}\left(1, \frac{1}{2}; 1-\ep; -p^2t^2\right) = \frac{1}{\left(1+t^2p^2\right)^{1/2}} \,
{}_{2}F_{1}\left(\frac{1}{2}, -\ep; 1-\ep; q(p)\right) \,,
\label{Sum.1.3}
\ee
where $q$ was defined in Eq.~(\ref{y}) of the main text, and in what follows we suppress the $p$-dependence of $q(p)$ for ease of notation. The last ${}_{2}F_{1}$-function admits the following representation,
\be
{}_{2}F_{1}\left(\frac{1}{2}, -\ep; 1-\ep; q\right) = \sum_{m=0}^{\infty} \, \frac{\Gamma(m+\frac{1}{2})}{m!\Gamma(\frac{1}{2})} \, \frac{-\ep}{m-\ep}  \, q^m
= 1- \ep \sum_{m=1}^{\infty} \, \frac{\Gamma(m+\frac{1}{2})}{m!\Gamma(\frac{1}{2})} \, \frac{1}{m-\ep}\, q^m \, .
\ee
Using the integral representation for the factor $1/(m-\ep) =\int_0^1 dz \, z^{m-1-\ep}$, we have,
\be
{}_{2}F_{1}\left(\frac{1}{2}, -\ep; 1-\ep; q\right) = 1- \ep \int_0^1 \frac{dz}{z^{1+\ep}} \, \left[\frac{1}{\sqrt{1-zq}} -1 \right] = 1- \ep \int_0^q \frac{dz_1 q^{\ep}}{z_1^{1+\ep}} \, \left[\frac{1}{\sqrt{1-z_1}} -1 \right].
\label{Sum.1.4}
\ee
Combining these results, Eq.~(\ref{Sum.1.2}) can be written as,
\be
\sum_{m=0}^{\infty} \, \frac{\Gamma(m+\frac{3}{2})}{\Gamma(m+2-\ep)}  \, {(-(tp)^2)}^m
= - \frac{\Gamma(\frac{1}{2})}{\Gamma(1-\ep)(tp)^2} \left[ \frac{1}{\left(1+t^2p^2\right)^{1/2}} \, \left\{
1- \ep J_{1}(p)
\right\} \, -1 \right] \, ,
\label{Sum.1.5}
\ee
where $J_1(p)$ is defined in Eq.~(\ref{Ji}). 

The final result for $F_{1}(t)$ reads,
\be
F_{1}(t) = - \frac{3(1-\ep)}{2(1+\ep)t^2} \, \hat{K}_1
\, \int_0^1 \, dp \, p^{\ep-1}(1-p)^{-1/2-\ep}
\, \left[ \frac{1}{\left(1+t^2p^2\right)^{1/2}} \, \left\{
1- \ep J_{1}(p)
\right\} \, -1 \right] \, ,
\label{J122.1.4}
\ee
where the normalization $\hat{K}_1$ was determined in Eq.~(\ref{Ki}). We note that the elliptic structure of Eq.~(\ref{J122.1.4}) is carried by the product $(1-p)^{-1/2-\ep} (1+t^2p^2)^{-1/2}$.

\subsection*{The second hypergeometric function}

The second ${}_4F_3$-hypergeometric function of $J_{1,2,2}$ admits the following series representation,
\bea
&&F_{2}(t) \equiv
 {}_{4}F_{3}\left(\left.
\begin{array}{c}
1,\frac{1}{2}+\ep,1+\frac{\ep}{2},1+\ep\\
\frac{3}{2}-\frac{\ep}{2},\frac{3}{4}+\frac{\ep}{2},\frac{5}{4}+\frac{\ep}{2}
\end{array}
\right| -t^2 \right)  \,
\nonumber \\
&&= \sum_{m=0}^{\infty}
\frac{\Gamma(m+\frac{1}{2}+\ep)\Gamma(m+1+\frac{\ep}{2})\Gamma(m+1+\ep)}{\Gamma(m+\frac{3}{2}-\frac{\ep}{2})
\Gamma(m+\frac{3}{4}+\frac{\ep}{2})\Gamma(m+\frac{5}{4}+\frac{\ep}{2})} \frac{\Gamma(\frac{3}{2}-\frac{\ep}{2})
\Gamma(\frac{3}{4}+\frac{\ep}{2})\Gamma(\frac{5}{4}+\frac{\ep}{2})}{\Gamma(\frac{1}{2}+\ep)\Gamma(1+\frac{\ep}{2})\Gamma(1+\ep)} {(-t^2)}^m.
\label{J122.2.1}
\eea
As in the previous section, we express $F_{2}(t)$ as,
\be
\sum_{m=0}^{\infty} \, \frac{\Gamma(m+1+\frac{\ep}{2})\Gamma(2m+2+2\ep)}{\Gamma(m+\frac{3}{2}-\frac{\ep}{2})\Gamma(2m+\frac{3}{2}+\ep)} \, \frac{\Gamma(\frac{3}{2}-\frac{\ep}{2})
\Gamma(\frac{1}{2}+\ep)}{\Gamma(1+\frac{\ep}{2})\Gamma(1+2\ep)}
\, {(-t^2)}^m  \, ,
\label{J122.2.2}
\ee
where we used the standard identity,
\be
{}_{2}F_{1}\Big(a,b;c;z\Bigr)= (1-z)^{c-b-a}  \,{}_{2}F_{1}\left(c-a,c-b;c;z\right) \, .
\label{2F1.2}
\ee
By turning to integral representations for gamma functions,
\be
\frac{\Gamma(2m+1+2\ep)}{\Gamma(2m+\frac{3}{2}+\ep)} = \int_0^1 \, dp \, \frac{p^{2m+2\ep}(1-p)^{-1/2-\ep}}{\Gamma(\frac{1}{2}-\ep)} \,,
\label{Int.p.2}
\ee
we find,
\be
   F_{2}(t) =
\int_0^1 \, dp \, \frac{p^{2\ep}(1-p)^{-1/2-\ep}}{\Gamma(\frac{1}{2}-\ep)}
\sum_{m=0}^{\infty} \, \frac{\Gamma(m+1+\frac{\ep}{2})}{\Gamma(m+\frac{3}{2}-\frac{\ep}{2})} \, \frac{\Gamma(\frac{3}{2}-\frac{\ep}{2})
\Gamma(\frac{3}{2}+\ep)}{\Gamma(+\frac{\ep}{2})\Gamma(1+2\ep)}
\, {(-(tp)^2)}^m   \, .
\label{J122.2.3}
\ee

As before, we consider first the $\ep=0$ limit,
\bea
&&\sum_{m=0}^{\infty} \, \frac{\Gamma(m+1)\Gamma(\frac{1}{2})}{\Gamma(m+\frac{3}{2})} \, \, {(-t^2)}^{m}
= 2 \, {}_{2}F_{1}\left(1,1; \frac{3}{2}; -t^2\right) = \frac{2}{\left(1+t^2\right)^{1/2}} \,
{}_{2}F_{1}\left(\frac{1}{2}, \frac{1}{2}; \frac{1}{2}; -t^2\right) \nonumber \\
&&= \frac{1}{\left(1+t^2\right)^{1/2}} \,
 \sum_{m=0}^{\infty} \, \frac{\Gamma(m+\frac{1}{2})}{m!\Gamma(\frac{1}{2})} \, \frac{1}{m+\frac{1}{2}} \, \, {(-t^2)}^{m}.
 \label{Sum.2.4}
\eea
By taking $1/(m+1/2)=\int^1_0 dz \, z^{m-1/2}$, we find ($z=s^2$),
\bea
&&\frac{1}{\left(1+t^2\right)^{1/2}} \,
 \sum_{m=0}^{\infty} \, \frac{\Gamma(m+\frac{1}{2})}{m!\Gamma(\frac{1}{2})} \, \frac{1}{m+\frac{1}{2}} \, \, {(-t^2)}^{m}
 = \frac{1}{\left(1+t^2\right)^{1/2}} \, \int^1_0 \frac{dz}{z} \, \frac{1}{\sqrt{1+zt^2}} \nonumber \\
&&= \frac{2}{\left(1+t^2\right)^{1/2}}  \int^1_0 \frac{d s}{\sqrt{1+s^2t^2}} = \frac{1}{t\sqrt{1+t^2}}  \,
\log \, \frac{\sqrt{1+t^2}+1}{\sqrt{1+t^2}-1}.
 \label{Sum.2.5}
\eea
In the general case,
\bea
&&\sum_{m=0}^{\infty} \, \frac{\Gamma(m+1+\frac{\ep}{2})\Gamma(\frac{3}{2}-\frac{\ep}{2})}{\Gamma(m+\frac{3}{2}-\frac{\ep}{2})
 \Gamma(1+\frac{\ep}{2})}
 \, \, {(-t^2)}^{m} \nonumber \\
&&=  {}_{2}F_{1}\left(1+\frac{\ep}{2},1; \frac{3}{2}-\frac{\ep}{2}; -t^2\right) = \frac{1}{\left(1+t^2\right)^{1/2+\ep}} \,
{}_{2}F_{1}\left(\frac{1}{2}-\ep, \frac{1}{2}-\frac{\ep}{2}; \frac{3}{2}-\frac{\ep}{2}; -t^2\right) \nonumber \\
&&= \frac{1}{\left(1+t^2\right)^{1/2+\ep}} \,
 \sum_{m=0}^{\infty} \, \frac{\Gamma(m+\frac{1}{2}-\ep)}{m!\Gamma(\frac{1}{2}-\ep)} \, \frac{(\frac{1}{2}-\frac{\ep}{2})}{m+\frac{1}{2}-\frac{\ep}{2}} \, \, {(-t^2)}^{m},
 \label{Sum.2.6}
\eea
and by taking $1/(m+1/2-\ep/2)=\int^1_0 dz \, z^{m-1/2-\ep/2}$, we have ($z=s^2$),
\bea
&&\frac{1-\ep}{2\left(1+t^2\right)^{1/2+\ep}} \,
 \sum_{m=0}^{\infty} \, \frac{\Gamma(m+\frac{1}{2}-\ep)}{m!\Gamma(\frac{1}{2}-\ep)} \, \frac{1}{m+\frac{1}{2}-\frac{\ep}{2}} \, \, {(-t^2)}^{m}
 = \frac{1-\ep}{2\left(1+t^2\right)^{1/2+\ep}} \, \int^1_0 \frac{dz}{z^{\frac{1}{2}+\frac{\ep}{2}}} \, \frac{1}{(1+zt^2)^{\frac{1}{2}-\ep}} \nonumber \\
 &&= \frac{1-\ep}{\left(1+t^2\right)^{1/2+\ep}}  \int^1_0 \frac{d s}{s^{\ep}} \, \frac{1}{(1+s^2t^2)^{\frac{1}{2}-\ep}}
 \equiv \frac{1-\ep}{\left(1+t^2\right)^{1/2+\ep} t^{1-\ep}} \, J_2(p) \, ,
 \label{Sum.2.7}
\eea
where $J_2(p)$ is defined in Eq.~(\ref{Ji}).
%

The final results for $ F_{2}(t)$ reads,
\be
F_{2}(t)=
\frac{(1-\ep)(1+2\ep)}{2^{2+\ep}t^{1-\ep}} \, \hat{K}_2 \,
\int_0^1 \, dp \, p^{3\ep}(1-p)^{-1/2-\ep} \,
\frac{1}{\left(1+t^2p^2\right)^{1/2+\ep}} \, J_{2}(p) \, ,
\label{J122.2.8}
\ee
where  the normalization $\hat{K}_1$ is defined in Eq.~(\ref{Ki}).

\subsection*{The third hypergeometric function}

The third ${}_4F_3$-hypergeometric function admits the following series representation,
\bea
&&F_{3}(t) \equiv
{}_{4}F_{3}\left(\left.
\begin{array}{c}
1,\frac{3}{2}+\frac{\ep}{2},1+\ep,\frac{3}{2}+\ep\\
2-\frac{\ep}{2},\frac{5}{4}+\frac{\ep}{2},\frac{7}{4}+\frac{\ep}{2}
\end{array}
\right| -t^2 \right) \,
\nonumber \\
&&= \sum_{m=0}^{\infty} \,
\frac{\Gamma(m+\frac{3}{2}+\frac{\ep}{2})\Gamma(m+\frac{3}{2}+\ep)\Gamma(m+1+\ep)}{\Gamma(m+2-\frac{\ep}{2})
\Gamma(m+\frac{5}{4}+\frac{\ep}{2})\Gamma(m+\frac{7}{4}+\frac{\ep}{2})} \frac{\Gamma(2-\frac{\ep}{2})
\Gamma(\frac{5}{4}+\frac{\ep}{2})\Gamma(\frac{7}{4}+\frac{\ep}{2})}{\Gamma(\frac{3}{2}+\frac{\ep}{2})
\Gamma(1+\ep)\Gamma(\frac{3}{2}+\ep)}
\, {(-t^2)}^m  .
\label{J122.3.1}
\eea
The previous sum simplifies to,
\be
\sum_{m=0}^{\infty} \, \frac{\Gamma(m+1+\frac{\ep}{2})\Gamma(2m+2+2\ep)}{\Gamma(m+\frac{3}{2}-\frac{\ep}{2})\Gamma(2m+\frac{3}{2}+\ep)} \, \frac{\Gamma(\frac{3}{2}-\frac{\ep}{2})
\Gamma(\frac{1}{2}+\ep)}{\Gamma(1+\frac{\ep}{2})\Gamma(1+2\ep)}
\, {(-t^2)}^m  \, .
\label{J122.3.2}
\ee
By considering the following representations for ratios of gamma functions, 
\be
\frac{\Gamma(2m+2+2\ep)}{\Gamma(2m+\frac{5}{2}+\ep)} = \int_0^1 \, dp \, \frac{p^{2m+1+2\ep}(1-p)^{-1/2-\ep}}{\Gamma(\frac{1}{2}-\ep)} \,,
\label{Int.p.3}
\ee
we find,
\be
   F_{3}(t) =
\int_0^1 \, dp \, \frac{p^{1+2\ep}(1-p)^{-1/2-\ep}}{\Gamma(\frac{1}{2}-\ep)}
\sum_{m=0}^{\infty} \, \frac{\Gamma(m+\frac{3}{2}+\frac{\ep}{2})}{\Gamma(m+2-\frac{\ep}{2})} \, \frac{\Gamma(2-\frac{\ep}{2})
\Gamma(\frac{5}{2}+\ep)}{\Gamma(\frac{3}{2}+\frac{\ep}{2})\Gamma(2+2\ep)}
\, {(-(tp)^2)}^m  \, .
\label{J122.3.3}
\ee

As in the previous sections, we consider first the limit of the right hand side of the previous equation, which is equivalent to Eq.~(\ref{Sum.1.1}). In the general case,
\bea
&&\sum_{m=0}^{\infty} \, \frac{\Gamma(m+\frac{3}{2}+\frac{\ep}{2})}{\Gamma(m+2-\frac{\ep}{2})}  \, {(-(tp)^2)}^m  =
\sum_{m=1}^{\infty} \, \frac{\Gamma(m+\frac{1}{2}+\frac{\ep}{2})}{\Gamma(m+1-\frac{\ep}{2})} \, {(-(tp)^2)}^{m-1} \nonumber \\
&&= - \frac{\Gamma(\frac{1}{2}+\frac{\ep}{2})}{\Gamma(1-\frac{\ep}{2})(tp)^2} \left[ {}_{2}F_{1}\left(1, \frac{1}{2}+\frac{\ep}{2}; 1-\frac{\ep}{2}; -p^2t^2\right)
 \, -1 \right].
\label{Sum.3.4}
\eea
By using Eq.~(\ref{2F1.1}) we obtain,
\be
{}_{2}F_{1}\left(1, \frac{1}{2}+\frac{\ep}{2}; 1-\frac{\ep}{2}; -p^2t^2\right) = \frac{1}{\left(1+t^2p^2\right)^{1/2+\frac{\ep}{2}}} \,
{}_{2}F_{1}\left(\frac{1}{2}+\frac{\ep}{2}, -\frac{\ep}{2}; 1-\frac{\ep}{2}; q\right) \, .
\label{Sum.3.5}
\ee
%
The last ${}_{2}F_{1}$-function admits the following representation,
\be
{}_{2}F_{1}\left(\frac{1}{2}+\frac{\ep}{2}, -\frac{\ep}{2}; 1-\frac{\ep}{2}; q\right) =
\sum_{m=0}^{\infty} \, \frac{\Gamma(m+\frac{1}{2}+\frac{\ep}{2})}{m!\Gamma(\frac{1}{2}+\frac{\ep}{2})} \,
\frac{(-\frac{\ep}{2})}{m-\frac{\ep}{2}}  \, q^m
= 1- \frac{\ep}{2} \sum_{m=1}^{\infty} \, \frac{\Gamma(m+\frac{1}{2}+\frac{\ep}{2})}{m!\Gamma(\frac{1}{2}+\frac{\ep}{2})} \, \frac{1}{m-\frac{\ep}{2}}\, q^m \, .
\label{Sum.3.6}
\ee
By using the integral representation $1/(m-\frac{\ep}{2}) =\int_0^1 dz \, z^{m-1-\frac{\ep}{2}}$, we find ($z_1=z q$),
\begin{align}
{}_{2}F_{1}\left(\frac{1}{2}+\frac{\ep}{2}, -\frac{\ep}{2}; 1-\frac{\ep}{2}; q\right) &= 1-\frac{\ep}{2} \int_0^1 \frac{dz}{z^{1+\frac{\ep}{2}}} \, \left[\frac{1}{(1-z q)^{\frac{1}{2}+\frac{\ep}{2}}} -1 \right] \nonumber\\
&= 1- \frac{\ep}{2} \int_0^q \frac{dz_1 q^{\frac{\ep}{2}}}{z_1^{1+\frac{\ep}{2}}} \, \left[\frac{1}{(1-z_1)^{\frac{1}{2}+\frac{\ep}{2}}} -1 \right],
\label{Sum.3.7}
\end{align}
so that Eq.~(\ref{Sum.1.2}) reads,
\be
\sum_{m=0}^{\infty} \, \frac{\Gamma(m+\frac{3}{2}+\frac{\ep}{2})}{\Gamma(m+2-\frac{\ep}{2})}  \, {(-(tp)^2)}^m
= - \frac{\Gamma(\frac{1}{2}+\frac{\ep}{2})}{\Gamma(1-\frac{\ep}{2})(tp)^2} \left[ \frac{1}{\left(1+t^2p^2\right)^{1/2+\frac{\ep}{2}}} \, \left\{
1- \frac{\ep}{2} J_{3}(p)
\right\} \, -1 \right] \, ,
\label{Sum.3.8}
\ee
where $J_3(p)$ is defined in Eq.~(\ref{Ji}). 

The final result for $ F_{3}(t)$ is,
\be F_{3}(t)=
-\frac{(1-\frac{\ep}{2})(\frac{3}{2}+\ep)}{(1+\ep)t^{2}} \, \hat{K}_2 \,
\int_0^1 \, dp \, p^{\ep}(1-p)^{-1/2-\ep} 
\, \left[ \frac{1}{\left(1+t^2p^2\right)^{1/2+\frac{\ep}{2}}} \, \left\{
1- \frac{\ep}{2} J_{3}(p)
\right\} \, -1 \right] \, .
\ee

\def\theequation{B\arabic{equation}}
\setcounter{equation}{0}

\section{Leading terms of the $\ep$-expansion and one-fold integrals }
\label{App:pole cancellation}
In this section we derive a one-fold integral representation for the first two orders of the $\epsilon$-expansion of $J_{1,2,2}$, Eq.~(\ref{J122}). This representation generalises the results of Refs. \cite{Kniehl:2005bc} and \cite{Kniehl:2019vwr}, where only the finite part of the expansion was considered. The first two $\epsilon$ orders considered here can be expressed as one-fold integrals over logarithms and dilogarithms with algebraic prefactors. A similar analysis shows that to arbitrary order of the dimensional regulator the result is in terms of one-fold integrals over higher weight MPLs. We leave this analysis for future work.

\subsection*{The inner integrals}

We start by considering integral $J_{2}(p)$, defined in Eq.~(\ref{Ji}), at order $\epsilon^0$,
\be
J_{2}(p,\ep =0)= \int^{tp}_0 \frac{ds}{\sqrt{1+s^2}} \, ,
\label{J2.0}
\ee
which can be evaluated directly by means of the variable change,
\be
s_2 = \frac{\sqrt{1+s^2}-s}{\sqrt{1+s^2}+s}\,,
  \label{s2}
\ee
leading to,
\be
J_{2}(p,\ep =0)= \frac{1}{2} \, \int^{1}_{R_2} \frac{ds_2}{s_2} = -  \frac{1}{2} \, \log R_2 \equiv J_{2,0}(p)\,,
\ee
with
\be
R_2 = \frac{\sqrt{1+tp^2}-tp}{\sqrt{1+tp^2}+tp} = \frac{1-\sqrt{q}}{1+\sqrt{q}} \, .
  \label{R2}
\ee
By means of the same variable change, we evaluate the next $\epsilon$ order,
\be
J_{2}(p)= \frac{1}{2^{1+\ep}} \, \int^{1}_{R_2} \frac{ds_2}{s_2^{1+\ep/2}} \, \frac{(1+s_2)^{2\ep}}{(1-s_2)^{2\ep}}
=  J_{2,0}(p) + \ep  J_{2,1}(p) +\mathcal{O}(\epsilon^2)\, ,
\label{J2.0a}
\ee
where $J_{2,0}(p)$ is given in (\ref{J2.0a}) and,
\be 
J_{2,1}(p)= \frac{1}{8} \, \log^2 R_2 + \zeta_2 + {\rm Li}_2(-R_2) - \frac{1}{2} \, {\rm Li}_2(R_2) \, .
\label{J21}
\ee
We now consider integrals $J_{1}(p)$ and $J_{3}(p)$, defined in Eq.~(\ref{Ji}), at order $\epsilon^0$,
\be
J_{1}(p,\ep =0)=J_{3}(p,\ep =0)= \int^{q}_0 \frac{dz}{z} \, \left(\frac{1}{\sqrt{1-z}}-1\right) \, .
\label{J13.0}
\ee
By introducing a regulator $\delta$ we have,
\be
\int^{y}_{\delta} \frac{dz}{z} = \log q-\log \delta \,,
\ee
while the remaining term can be evaluated by the variable change,
\begin{equation}
z=1-s^2, \quad s=\frac{(1-s_1)}{(1+s_1)}\,,    
\end{equation}
and,
\be
\int^{q}_{\delta} \frac{dz}{z\sqrt{1-z}}  = \int^{R_1}_{\delta/4} \frac{ds_1}{s_1} = \log R_1-\log \frac{\delta}{4} \,,\quad 
R_1 = \frac{1-\sqrt{1-q}}{1+\sqrt{1-q}} \,.
  \label{R1}
  \ee
The full result can be written as,
\be
J_{i}(p,\ep =0)= \log (4R_1)-\log \delta = \log \frac{4R_1}{q} \equiv J_{i,0}(p), ~~~ (i=1,3)\, .
\label{J13.0a}
\ee
At the next order we have,
\be
J_{i}(p)
=  J_{i,0}(p) + \ep  J_{i,1}(p) +\mathcal{O}(\epsilon^2),   ~~~ (i=1,3)\, ,
\label{Ji.0a}
\ee
where $J_{1,0}(p)=J_{3,0}(p)$ are given in (\ref{J13.0a}) and,
\be 
J_{1,1}(p)= \overline{J}_{1,1}(p)-
2{\rm Li}_2(-R_1),~~ J_{1,1}(p)= \frac{1}{2} \, \overline{J}_{1,1}(p)+2{\rm Li}_2(R_1)-4 {\rm Li}_2(-R_1) \, ,
\label{J13}
\ee
with
\be
\overline{J}_{1,1}(p) =  \log q \log (4R_1) - \frac{1}{2} \, \log^2 q - \log 4 \log R_1 \, .
\label{oJ11}
\ee

\subsection*{Result for the sunrise integral}

Combining Eqs.~(\ref{Ji122N1}) and~(\ref{Ki}) to the right hand side of Eq.~(\ref{J122N}), we obtain the following expression for $J_{1,2,2}$,
\be
J_{1,2,2} = \hat{N}_1 \, \frac{\hat{K}_1}{4\ep t^2} \, \hat{J}_{1,2,2},~~
\hat{J}_{1,2,2} = \biggl[-\frac{1}{\ep} I^{(1)}(t) + {\left(\frac{t^2}{2}\right)}^{\ep} \,
  I^{(2)}(t) + \frac{1}{\ep} \, {\left(\frac{t}{2}\right)}^{\ep} \, I^{(3)}(t) \biggr] \, .
\label{J122a}
\ee
In order to obtain the $\ep$-expansion of $J_{1,2,2}$ up to and including $\mathcal{O}(\epsilon)$, we use the expressions for integrals $I^{(i)}(t)$ $(i=1,2,3)$ of Eq.~(\ref{I1,2}) and their integral representation Eq.~(\ref{eq:2integrals p}). We have,
\bea
&&I_{1}^{(1)}(t) = \int_0^1 \frac{dp}{p\sqrt{1-p}} \, \Bigg(\frac{1}{\sqrt{1+p^2t^2}}-1\Bigg) \,
\left[1+ \ep \, l_1 + \frac{\ep^2}{2} \, l_1^2 \right]+\mathcal{O}(\epsilon^2)\,, \nonumber \\
&&I_{2}^{(1)}(t) = \int_0^1 \frac{dp}{p\sqrt{1-p}} \, \frac{1}{\sqrt{1+p^2t^2}} \,
\left[J_{10}+ \ep \, (l_1J_{10} + J_{11}) \right]+\mathcal{O}(\epsilon^2), \nonumber \\
&&I^{(2)}(t) = \int_0^1 \frac{dp}{p\sqrt{1-p}} \, \frac{1}{\sqrt{1+p^2t^2}} \,
  \left[J_{20}+ \ep \, (l_2J_{20} + J_{21}) \right]+\mathcal{O}(\epsilon^2)\,,
  \nonumber \\
&&I_{1}^{(3)}(t) = \int_0^1 \frac{dp}{p\sqrt{1-p}} \,\left( \frac{1}{\sqrt{1+p^2t^2}} \,
\left[1+ \ep \, l_{32} + \frac{\ep^2}{2} \, l_{32}^2 \right] - \left[1+ \ep \, l_{31} + \frac{\ep^2}{2} \, l_{31}^2 \right]
\right) +\mathcal{O}(\epsilon^2)\, ,\nonumber \\
&&I_{2}^{(3)}(t) = \int_0^1 \frac{dp}{p\sqrt{1-p}} \, \frac{1}{\sqrt{1+p^2t^2}} \,
\left[J_{30}+ \ep \, (l_{32}J_{30} + J_{31}) \right]+\mathcal{O}(\epsilon^2)\,,
\label{Iiep}
\eea
where,
\bea
&&l_1=\log \left(\frac{p}{1-p}\right),~~l_2=\log \left(\frac{p^3}{(1-p)(1+p^2t^2)}\right),~~ \nonumber \\
&&l_{31}=\log \left(\frac{p^2}{1-p}\right),~~l_{32}=\log \left(\frac{p}{(1-p)\sqrt{1+p^2t^2}}\right) ,\, 
\label{li}
\eea
with $J_{i0}$ and $J_{i1}$ $(i=1,2,3)$ given in Eqs.~(\ref{J2.0a}), (\ref{J21}),
(\ref{J13.0a})-(\ref{oJ11}).


Combining all terms, we obtain the following finite expression,
\begin{align}
\hat{J}_{1,2,2}& =  \int_0^1 \frac{dp}{p\sqrt{1-p}} \left( \frac{1}{2\sqrt{1+p^2t^2}}
\log (R_1R_2) -
\log\left(\frac{tp}{2}\right) + \ep \Bigg[\frac{1}{\sqrt{1+p^2t^2}} \Phi(q) - \log\left(\frac{tp^3}{2(1-p)^2}\right)
\Bigg] \right)\nonumber\\
&+\mathcal{O}(\epsilon^2),
\label{hJ122}
\end{align}
where,
\be
\Phi(q)= \frac{3}{2} \, \log^2 2 +  \frac{1}{2} \, \log\left(\frac{p\,q}{4(1-p)}\right) \, \log (R_1R_2) -\zeta_2 -
\Phi_1(R_1)-\Phi_2(R_2),
\label{Phi}
\ee
and,
\bea
&&\Phi_1(R_1)= \frac{1}{2} \, \log 2 \, \log R_1 + {\rm Li}_2(-R_1) +  \frac{3}{8} \,\log^2 R_1 , \nonumber \\
&&\Phi_2(R_2)= \frac{1}{8} \, \log^2 R_2 + {\rm Li}_2(-R_2) -  \frac{1}{2} \,  {\rm Li}_2(R_2) \, .
\label{Phi_i}
\eea


\def\theequation{C\arabic{equation}}
\setcounter{equation}{0}

\section{Definitions for elliptic polylogarithms}
\label{app:definitions}

In this section we provide the definitions for the various factors appearing in Eq.~(\ref{main}). Specifically, the eMPLs expressions for the relevant logarithms are defined as,
\begin{align}
    L_1&=\Ef{1}{-1}{x}-2 \Ef{1}{0}{x}+\Ef{1}{1}{x},\nonumber\\
    L_2&=\sum_{i=1}^4\Ef{1}{a_i}{x}-2 \Ef{1}{-1}{x}-2 \Ef{1}{1}{x}+\log \left(t^2+1\right)-2 \log (t),\nonumber\\
    L_3&=-\sum_{i=1}^4\Ef{1}{a_i}{x}+3 \Ef{1}{-1}{x}-2 \Ef{1}{0}{x}+3 \Ef{1}{1}{x}-\log \left(t^2+1\right)+2 \log (t),\nonumber\\
    L_4&=\sum_{i=1}^4\Ef{1}{a_i}{x}-\Ef{1}{-1}{x}-\Ef{1}{1}{x}+\log \left(t^2+1\right)-\log (t),\nonumber\\
    L_5&=-\sum_{i=1}^4\Ef{1}{a_i}{x}+3 \Ef{1}{-1}{x}-2 \Ef{1}{0}{x}+3 \Ef{1}{1}{x}-\log \left(t^2+1\right),\nonumber\\
    L_6&=2 \Ef{1}{-1}{x}-2 \Ef{1}{0}{x}+2 \Ef{1}{1}{x},\nonumber\\
    L_7&=-\frac{1}{2} \sum_{i=1}^4\Ef{1}{a_i}{x}+2 \Ef{1}{-1}{x}-2 \Ef{1}{0}{x}+2 \Ef{1}{1}{x}-\frac{1}{2} \log \left(t^2+1\right),\nonumber\\
    L_8&=-\sum_{i=1}^4\Ef{1}{a_i}{x}+3 \Ef{1}{-1}{x}-2 \Ef{1}{0}{x}+3 \Ef{1}{1}{x}-\log \left(t^2+1\right)+\log (t),\nonumber\\
    L_9&=\frac{1}{2}\sum_{i=1}^4 \Ef{1}{a_i}{x}-\Ef{1}{-1}{x}-\Ef{1}{1}{x}+\frac{1}{2} \log \left(t^2+1\right)-\log (t).
\end{align}

while the primitives of the relevant integration kernels are defined as,

\begin{align}
    K_1&=\Ef{-1}{-1}{x}-\Ef{-1}{1}{x}-\Ef{1}{-1}{x},\nonumber\\
    K_2&=\Ef{-1}{-1}{x}-\Ef{-1}{1}{x}-\Ef{1}{1}{x},\nonumber\\
    K_3&=-\sum_{i=1}^4\Ef{1}{a_i}{x}+2 \Ef{-1}{-1}{x}+2 \Ef{-1}{1}{x}-2 \Ef{1}{-1}{x},\nonumber\\
    K_4&=\Ef{-1}{-1}{x}-\Ef{-1}{1}{x}-\Ef{1}{1}{x},\nonumber\\
    K_5&=-2 \Ef{-1}{\infty }{x},\nonumber\\
    K_6&=\Ef{1}{1}{x}-\Ef{1}{-1}{x},\nonumber\\
    K_7&=\Ef{-1}{-1}{x}-\Ef{-1}{1}{x}-\Ef{1}{1}{x},\nonumber\\
    K_8&=-\sum_{i=1}^4\Ef{1}{a_i}{x}-2 \Ef{1}{-1}{x}-2 \Ef{1}{1}{x},\nonumber\\
    K_9&=2 \Ef{-1}{-1}{x}+2 \Ef{-1}{1}{x}+2 \Ef{1}{1}{x}.
\end{align}


\bibliographystyle{unsrt}

\bibliography{EllipK.bib,llipWW.bib,references.bib}

\end{document}